\documentclass{cimento}
\usepackage{graphicx}
\usepackage{cite}
\usepackage{amsmath}
\usepackage{epstopdf}
\usepackage{amssymb}

\def\h0 {\rm H_{0} }

\def\kms{\ifmmode\,{\rm km}\,{\rm s}^{-1}\else km$\,$s$^{-1}$\fi}
\def\hmpc{\ifmmode\,{\it h }^{-1}\,{\rm Mpc }\else $h^{-1}\,$Mpc\,\fi}

\def\\{\hfill\break}

\def\eg{{\it e.g.}}
\def\ie{{\it i.e.}}
\newcommand{\be}{\begin{equation}}
\newcommand{\ee}{\end{equation}}
\newcommand{\ba}{\begin{eqnarray}}
\newcommand{\ea}{\end{eqnarray}}
\newcommand{\brr}{\begin{array}}
\newcommand{\err}{\end{array}}
\newcommand{\bc}{\begin{center}}
\newcommand{\ec}{\end{center}}

\newcommand{\apj}{ApJ}

\newcommand{\apjl}{ApJ Lett.}

\newcommand{\pra}{Phys. Rev. A}

\newcommand{\MNRAS}{MNRAS}
\newcommand{\mnras}{MNRAS}

\title{Resolving the universe with multifractals}
\author{Cristiano G. Sabiu\from{ins:x},
Lu\`{i}s F. A. Teodoro\from{ins:y},
Martin A. Hendry\from{ins:y}}
\instlist{\inst{ins:x} Institute of Cosmology \& Gravitation, University of Portsmouth, England
  \inst{ins:y} Astronomy \& Astrophysics Group, University of Glasgow, Scotland}

\PACSes{\PACSit{98.80.Es}{Observational cosmology} 
              \PACSit{02.70.Rr}{General statistical methods} 
			 } 

\begin{document}

\maketitle
\footnotetext[1]{email: cris.sabiu@port.ac.uk}
\begin{abstract}
We present a new method for dealing with geometrical selection effects in galaxy surveys while using a multifractal framework. 
The power of multifractal analysis lies in its connection to higher order moments, in that it not only probes clustering on different 
scales but also different densities. Therefore any incompleteness issues must be correctly addressed before blindly applying this 
technique to real survey data.
\end{abstract}

\section{Introduction}
There has been much work done in trying to
characterise the galaxy distribution using multifractals\cite{Jones:1988,martinez+:1990,PC:2000}.
 One of the main problems which arises in a multifractal analysis is how to deal with observational selection
effects, \ie~`masks' in the survey region and a geometric boundary to the
survey itself. In this paper we will introduce a {\em volume\/} boundary correction which is 
rather similar to the approach developed by \cite{PC:2002}.

In \S \ref{MF} the methodology is constructed to analyse data within a 
multifractal framework, this then leads us to several problems in galactic surveys 
which must be accounted for. We will then discuss previous attempts at correcting 
for edge effects in \S\ref{BC} and show that there is still room for improvement. 
Then a  comparison is drawn between these different methods using a toy 
model in \S\ref{test}.

\section{Multifractal Formalism}\label{MF}
In this analysis we will adopt the procedure layed out in \cite{hentschel+:1983} 
to determine the R\'{e}nyi (Generalised) dimensions of a
point set embedded in a three-dimensional Euclidean space. The
probability of a galaxy, $j$, being within a sphere of radius $r$
centred on galaxy $i$ is,
\begin{eqnarray}
\label{prob}
{p}_{i}(r) = \displaystyle\frac{{n}_{i}(r)}{N} = \frac{1}{N}\displaystyle\sum_{j=1}^{N}\Phi(|r_{i}-r_{j}|-r).
\end{eqnarray}
Here $n_{i}(r)$ is the number of galaxies within radius $r$, $N$ is the total number 
of galaxies and $\Phi(x)$ is the Heaviside step function. Equation (\ref{prob}) can 
then be related to the partition sum via the correlation algorithm introduced in \cite{grassberger+:1983}
\begin{equation}
Z(q,r)=\frac{1}{M} \sum_{i=1}^{M}[{p}_{i}(r)]^{q-1} \propto r^{\tau (q)}.
\label{eq:partition_function}
\end{equation}
In this case $M$ is the number of counting spheres and $q$ defines the
generalised dimension we are investigating. $\tau (q)$ is the
scaling exponent, which is then related to the infinite set of
dimensions through,
\begin{equation}
\label{eq:dimension}
D_{q}={{\tau (q)} \over {q-1}}
\end{equation}

\section{Boundary Corrections}\label{BC}
Firstly, and most obviously, there is the {\em deflation} method.
This simply makes no correction at all, it only allows counting
spheres which lie completely within the survey. This dramatically
reduces the distance out to which our estimators can probe and also
reduces the number of large spheres we can average over, leading to
a sort of cosmic variance. 

The {\em deflation} method therefore wastes a great amount of data 
especially when used with a `pencil beam' or `thin fan' survey. It 
would be more useful if our estimators could sample all of the 
survey without worrying about its geometry. This however leads to 
counting cells exceeding the survey boundaries and consequently missing 
galaxies \cite{Hatton:1999}.

To deal with partially filled/empty cells we could use the 
{\em capacity} correction. This method populates the missing regions with 
galaxies (R2 in Fig.1). At first this may seem
like an appropriate coarse of action but we are then forced to
decide on the right distribution of galaxies to fill the void.
Previous attempts at analysing redshift surveys have used number densities like, $n\propto{r^3}$ 
(\eg \cite{Borgani+:1994}). This is of coarse assuming
an answer to the question we pose, as the exponent is fixed to be 3.
It is then not surprising that a transition to homogeneity has been
reported by groups using this approach.

In Fig.1 the counting sphere has
exceeded the geometrical boundary of the survey. The number of
galaxies counted in the sphere of radius $r$ is depleted which leads
to $p_{i}(r)$ being reduced through Eq. (\ref{prob}). To solve this
problem we could either add galaxies to the missing region like the
{\em capacity} correction or we could somehow modify the volume (hence the
name!). We can cast expression  (\ref{prob}) as,
\begin{eqnarray}
{p}_{i}(r) = \frac{\displaystyle{V}_{i}(r)\rho^{\star}(r)}{\displaystyle{N}} = \frac{{V}_{i}(r)} {{V}_{i}^{\star}(r)}.\frac{n_{i}^{\star}(r)}{N}\label{volcorr},
\end{eqnarray}
where now we have $V$ being the real volume of the sphere and
 $V^{\star}$ is the reduced volume. We have also introduced the reduced 
density as an intermediary step which need not be calculated and
related this with a reduced volume and number count,
$n_{i}^{\star}$. On its own this method can be visualised in Fig.1, 
as giving the missing region 2 the same density as region 1.
This would be wrong if density varies with distance, so that 
$\rho_{R1}\neq \rho_{R2}$. To overcome this problem we
assume {\em only} that the density does not vary with $\theta$ or
$\phi$ i.e. the universe is isotropic and hence Eq. (\ref{volcorr}) will
hold for fixed $r$. So to apply this method to a galaxy survey we must
count in spherical shells, correcting as we go and then integrate up
the shells at the end. This method is shown in Fig.2. 
The shells are individually corrected and summed according to,
\be
p_{i}(r)=\sum^{r}_{r=0}\alpha_{i}(r)\frac{n_{i}^{\star}}{N}
\ee
Where $\alpha_{i}(r)\equiv \ {V}/V^{\star}$, this is the
enhancement factor of the $i^{th}$ shell at radius $r$ and has value
$\geq$ 1. The main advantage of this method is that makes full use of the data. 
The deflation method on the other hand throws away a lot of useful data 
and hence increase the errors in their results.

\section{TESTING THE CORRECTIONS}\label{test}
In this section we will begin by applying the different corrections 
to a toy fractal model. This model can be used to compare and contrast the 
differing methods since it has an analytically determined dimension. We
also set up the fractal distributions to mimic real galaxy survey 
by being of comparable size (volume and number count). Then the 
{\em volume} correction will be used to analyse in detail the distribution of 
particles produced from $N$-body simulations. 

\subsection{MULTIPLICATIVE CASCADE}\label{fractal2}
To test further the power of our multifractal analysis we construct a hierarchical clustering model in two dimensions. 
This model is called the Multiplicative Cascade and was first discussed in Meakin (1987)\cite{1987PhRvA..36.2833M}, it has the advantage over 
the Levy Flight in that the whole of it Dq curve can be calculated analytically. 

The fractal is constructed as follows: The space is split into four equal parts, each part is then assigned a probability 
from the set $\lbrace p_1,p_2,p_3,p_4 \rbrace$ without replacement. Where $p_i \in [0,1]$. Each subspace is then 
divided again and assigned probabilities randomly from the same set and this is continued to the $N^{th}$ level.

At the $N^{th}$ level the probability of a cell being occupied is the product of the cell's $p_i$ and its parents and 
ancestors up to level 1 \ie~all the cells above it. In constructing this model down to level 8 we produce a $4^8$ 
array of cells each with its own probability. To then place particle in the space we invoke a Monte Carlo rejection scheme. 
Choosing x and y coordinates randomly we simply test if a random number between 0 and 1 is less or greater than the cell 
probability. We typically dust the probability density field with 5,000 points in a space of 256 x 256 \hmpc. The density field 
can be visualized in Fig.\ref{density}.

It can be shown \cite{Jones:1988} that as $L \rightarrow \infty$,
\be
D_q=\frac{\log_2\left( f^q_1+f^q_2+f^q_3+f^q_4\right)}{1-q},
\ee
where
\be
f_i=\frac{p_i}{\sum_i p_i}.
\ee
Fig.\ref{fig:3fractals} shows three fractal models with their measured and theoretical Renyi Dimensions.

\begin{table}
\centering
\begin{tabular}{r c c c c}
\hline \hline
 \ Model \ & \ $p_1$  & $p_2$ & $p_3$ & $p_4$ \\
\hline 
\ I \ &  1 & 1 & 1 & 0\\
\ II \ & 1 & 0.75 & 0.75 & 0.5\\
\ III \ &  1 & 0.5 & 0.5 & 0.25\\
\hline \hline
\end{tabular}
\caption{Parameters of the Multiplicative Cascade models}
\end{table}

\section{Conclusions \& Future Work}
Our `volume' correction and implementation of the Multifractal analysis successfully recovers the 
underlying Generalised dimensions of the multifractal distributions. It does a better job at correcting for the 
survey boundary than the methods discussed earlier and has the potential to work with much more complicated 
incomplete geometries. We hope soon to apply this method of analysis to real redshift surveys with the hope of 
uncovering the true hierarchical nature of the galaxy distribution. For a more detailed discusison see Sabiu 2007 \cite{sabiu:2007}.

\acknowledgments
LFAT is a Leverhulme Trust Research Fellow at the University of Glasgow. CGS is supported by a PhD studentship from STFC. 
CGS would like to thank Ben Hoyle for his help in constructing the Multiplicative Cascade. The authors would like to acknowledge 
Jun Pan for helpful duscussions. 


\newpage

\begin{figure}
\begin{minipage}[b]{2.5in}
\centering 
\includegraphics[scale=.3]{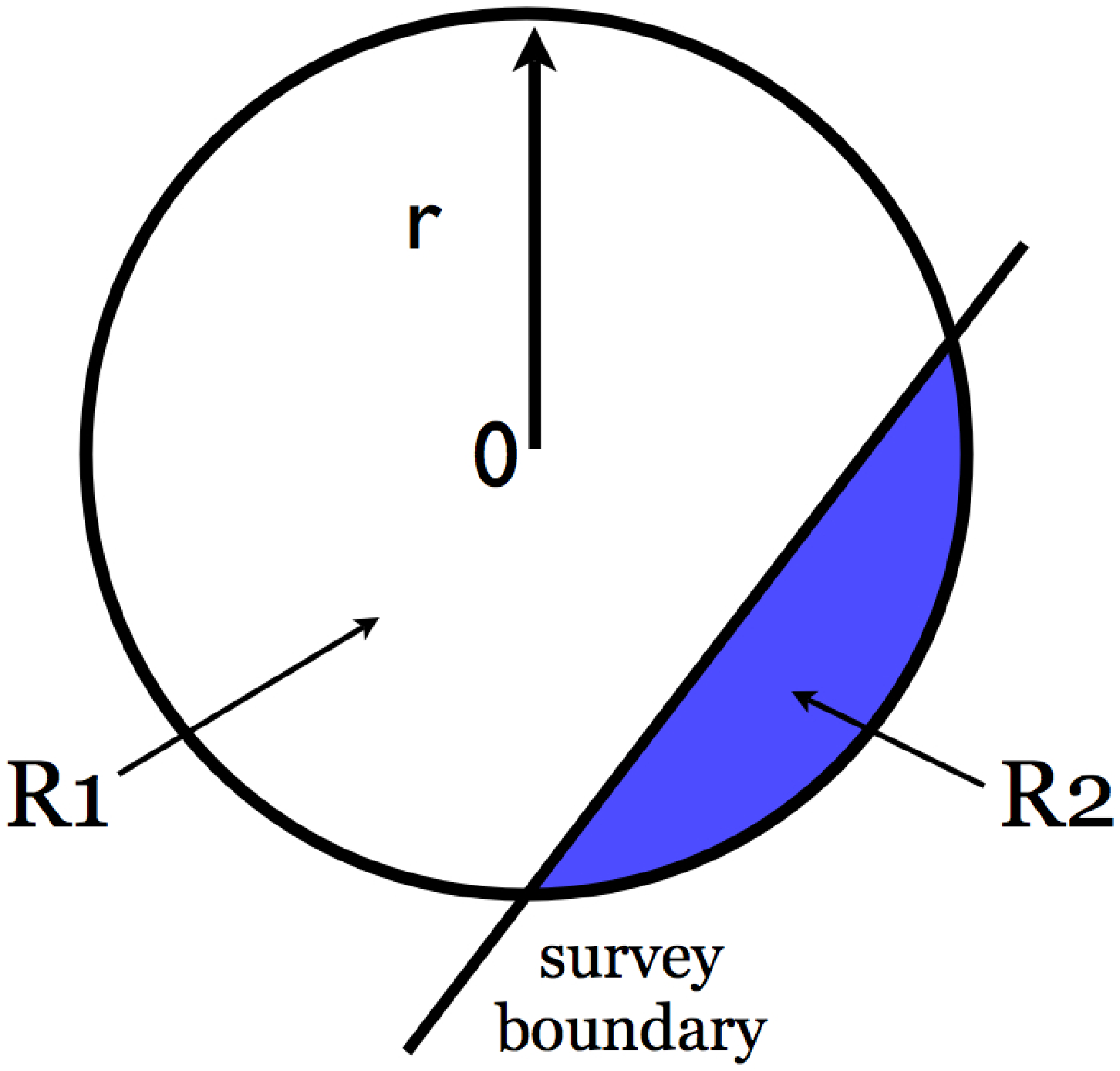}
\label{fig:circle1}
\caption{A counting sphere centred on galaxy O with radius r. Regions 
R1 \& R2 are inside and outside the survey respectively.}
\end{minipage}
\hfill
\begin{minipage}[b]{2.5in}
\centering 
\includegraphics[scale=.3]{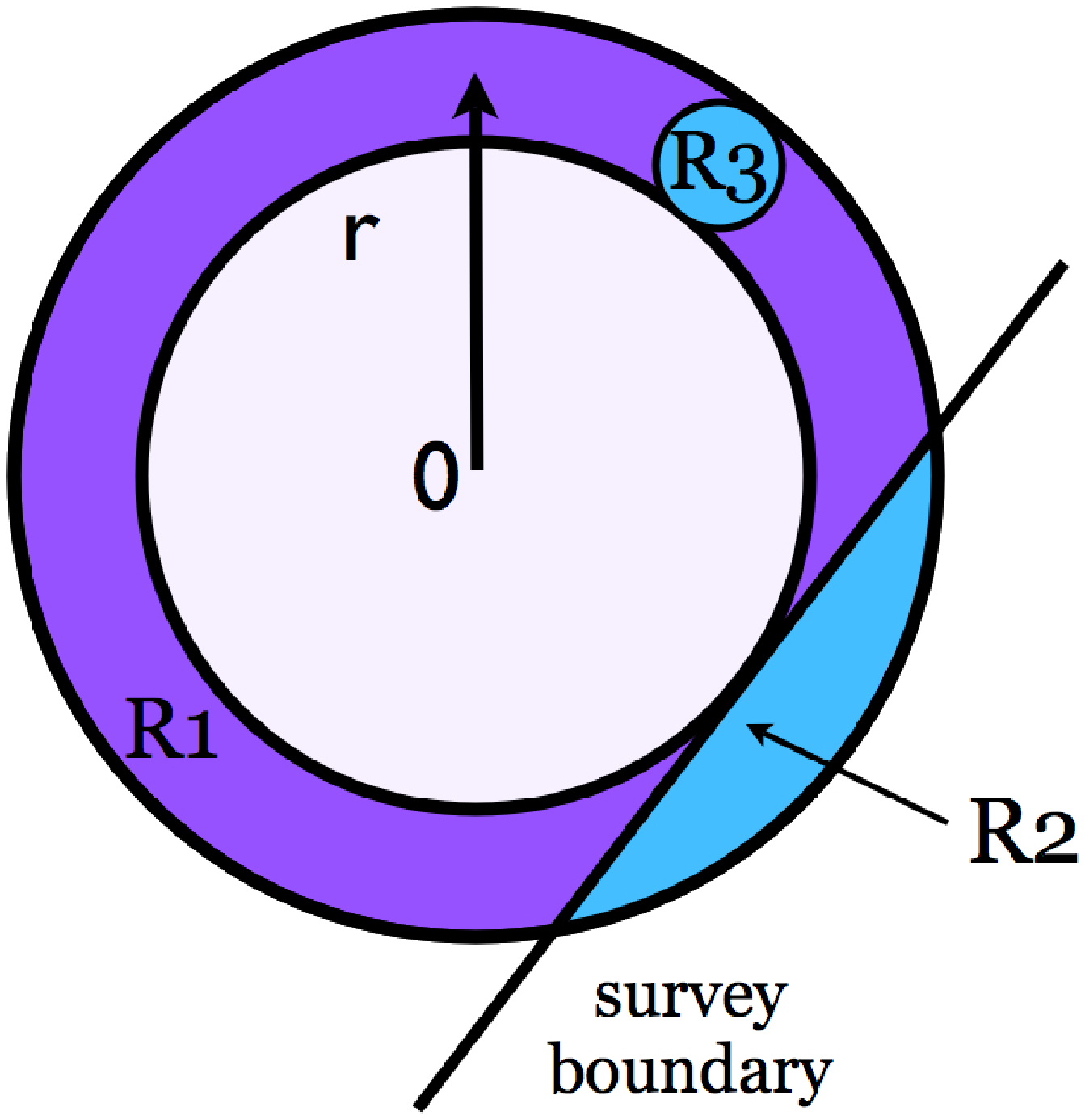}
\label{fig:circle}
\caption{ Region R1 is a shell inside the survey, R3 is a unobserved hole \& R2 is the boundary of the survey.}
\end{minipage}
\end{figure}

\begin{figure}
\centering
\includegraphics{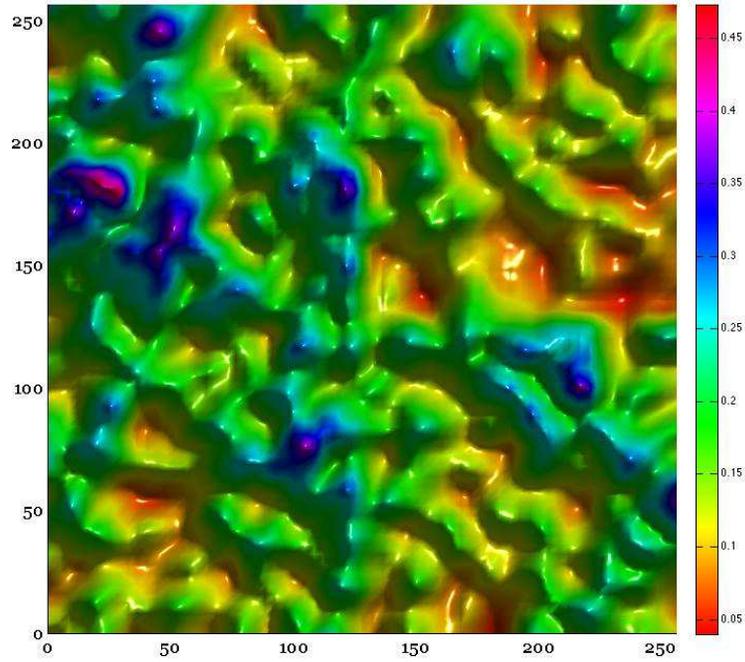}
\caption{\small This is an example of our probability density field as explained in \S\ref{fractal2}. }
\label{density}
\end{figure}

\begin{figure}%
\begin{minipage}[b]{2.5in}\centering \includegraphics[angle=90, width=2.8in]{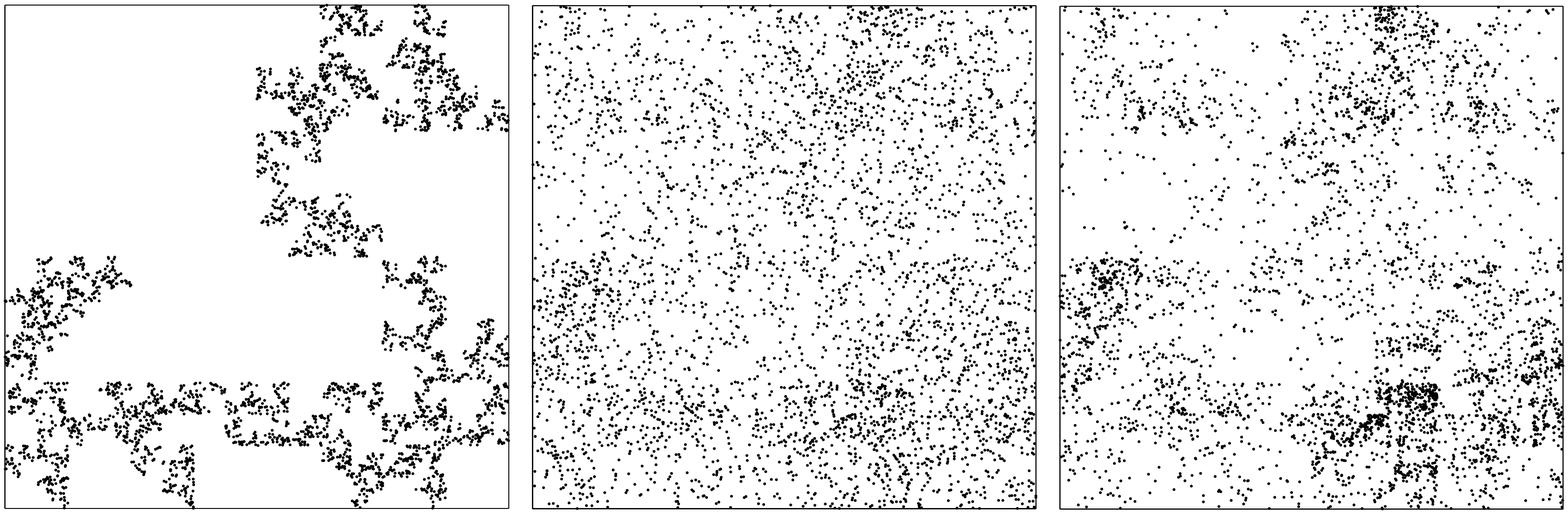}\end{minipage}%
\hfill%
\begin{minipage}[b]{2.5in}\centering \includegraphics[angle=90, width=3.3in]{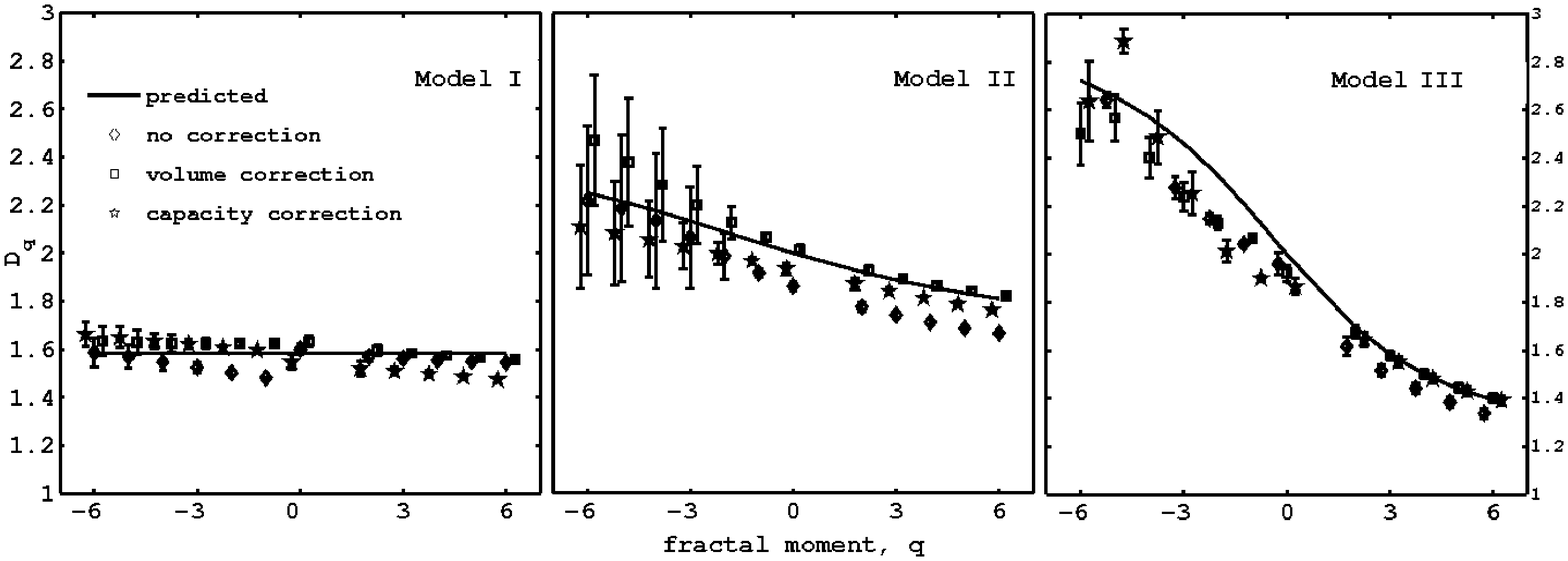}\end{minipage}%
\caption{\small Three realizations of our model fractal. Model I is a monofractal with uniform dimension $\log(3)/\log(2)$. Models II and III are multifractal distributions and are statistically inhomogeneous. The solid line is the theoretical model for the $D_q$ curve. }
\label{fig:3fractals}
\end{figure}

\end{document}